\input harvmac.tex

\input epsf.tex

\def\figin{\epsfcheck\figin}\def\figins{\epsfcheck\figins}
\def\epsfcheck{\ifx\epsfbox\UnDeFiNeD
\message{(NO epsf.tex, FIGURES WILL BE IGNORED)}
\gdef\figin##1{\vskip2in}\gdef\figins##1{\hskip.5in}
\else\message{(FIGURES WILL BE INCLUDED)}%
\gdef\figin##1{##1}\gdef\figins##1{##1}\fi}
\def\DefWarn#1{}
\def\figinsert{\goodbreak\midinsert}
\def\ifig#1#2#3{\DefWarn#1\xdef#1{fig.~\the\figno}
\writedef{#1\leftbracket fig.\noexpand~\the\figno}%
\figinsert\figin{\centerline{#3}}\medskip\centerline{\vbox{\baselineskip12pt
\advance\hsize by -1truein\noindent\footnotefont{\bf
Fig.~\the\figno:} #2}}
\bigskip\endinsert\global\advance\figno by1}
\def\frac#1#2{{#1 \over #2}}


\lref\amf{
  L.~F.~Alday and J.~M.~Maldacena,
  ``Gluon scattering amplitudes at strong coupling,''
  JHEP {\bf 0706}, 064 (2007)
  [arXiv:0705.0303 [hep-th]].
}
\lref\dualconf{
  J.~M.~Drummond, J.~Henn, V.~A.~Smirnov and E.~Sokatchev,
  ``Magic identities for conformal four-point integrals,''
  JHEP {\bf 0701}, 064 (2007)
  [arXiv:hep-th/0607160].
}
\lref\ams{
  L.~F.~Alday and J.~Maldacena,
  ``Comments on gluon scattering amplitudes via AdS/CFT,''
  JHEP {\bf 0711}, 068 (2007)
  [arXiv:0710.1060 [hep-th]].
}
\lref\grossmanes{
  D.~J.~Gross and J.~L.~Manes,
  ``The high-energy behavior of open string scattering''
  Nucl.\ Phys.\  B {\bf 326}, 73 (1989).
}
\lref\BernZX{
  Z.~Bern, L.~J.~Dixon, D.~C.~Dunbar and D.~A.~Kosower,
  ``One loop n point gauge theory amplitudes, unitarity and collinear limits,''
  Nucl.\ Phys.\  B {\bf 425}, 217 (1994)
  [arXiv:hep-ph/9403226].
}

\lref\gkp{
  S.~S.~Gubser, I.~R.~Klebanov and A.~M.~Polyakov,
  ``A semi-classical limit of the gauge/string correspondence,''
  Nucl.\ Phys.\  B {\bf 636}, 99 (2002)
  [arXiv:hep-th/0204051].
}

\lref\bds{
  Z.~Bern, L.~J.~Dixon and V.~A.~Smirnov,
  ``Iteration of planar amplitudes in maximally supersymmetric Yang-Mills
  theory at three loops and beyond,''
  Phys.\ Rev.\  D {\bf 72}, 085001 (2005)
  [arXiv:hep-th/0505205].
}

\lref\AnastasiouKJ{
  C.~Anastasiou, Z.~Bern, L.~J.~Dixon and D.~A.~Kosower,
  ``Planar amplitudes in maximally supersymmetric Yang-Mills theory,''
  Phys.\ Rev.\ Lett.\  {\bf 91}, 251602 (2003)
  [arXiv:hep-th/0309040].
}

\lref\MaldacenaRE{
  J.~M.~Maldacena,
  ``The large N limit of superconformal field theories and supergravity,''
  Adv.\ Theor.\ Math.\ Phys.\  {\bf 2}, 231 (1998)
  [Int.\ J.\ Theor.\ Phys.\  {\bf 38}, 1113 (1999)]
  [arXiv:hep-th/9711200].
}

\lref\GrossKZA{
  D.~J.~Gross and P.~F.~Mende,
  ``The High-Energy Behavior of String Scattering Amplitudes,''
  Phys.\ Lett.\  B {\bf 197}, 129 (1987).
}

\lref\BuchbinderHM{
  E.~I.~Buchbinder,
  ``Infrared Limit of Gluon Amplitudes at Strong Coupling,''
  Phys.\ Lett.\  B {\bf 654}, 46 (2007)
  [arXiv:0706.2015 [hep-th]].
}

\lref\KomargodskiWA{
  Z.~Komargodski,
  ``On collinear factorization of Wilson loops and MHV amplitudes in N=4 SYM,''
  arXiv:0801.3274 [hep-th].
}

\lref\BernAQ{
  Z.~Bern and D.~A.~Kosower,
  ``The Computation of loop amplitudes in gauge theories,''
  Nucl.\ Phys.\  B {\bf 379}, 451 (1992).
}

\lref\Druone{
  J.~M.~Drummond, G.~P.~Korchemsky and E.~Sokatchev,
  ``Conformal properties of four-gluon planar amplitudes and Wilson loops,''
  Nucl.\ Phys.\  B {\bf 795}, 385 (2008)
  [arXiv:0707.0243 [hep-th]].
}

\lref\BrandhuberYX{
  A.~Brandhuber, P.~Heslop and G.~Travaglini,
  ``MHV Amplitudes in N=4 Super Yang-Mills and Wilson Loops,''
  Nucl.\ Phys.\  B {\bf 794}, 231 (2008)
  [arXiv:0707.1153 [hep-th]].
}

\lref\Drutwo{
  J.~M.~Drummond, J.~Henn, G.~P.~Korchemsky and E.~Sokatchev,
  ``On planar gluon amplitudes/Wilson loops duality,''
  Nucl.\ Phys.\  B {\bf 795}, 52 (2008)
  [arXiv:0709.2368 [hep-th]].
}

\lref\Druthree{
  J.~M.~Drummond, J.~Henn, G.~P.~Korchemsky and E.~Sokatchev,
  ``Conformal Ward identities for Wilson loops and a test of the duality with
  gluon amplitudes,''
  arXiv:0712.1223 [hep-th].
}

\lref\BernEW{
  Z.~Bern, M.~Czakon, L.~J.~Dixon, D.~A.~Kosower and V.~A.~Smirnov,
  ``The Four-Loop Planar Amplitude and Cusp Anomalous Dimension in Maximally
  Supersymmetric Yang-Mills Theory,''
  Phys.\ Rev.\  D {\bf 75}, 085010 (2007)
  [arXiv:hep-th/0610248].
}

\lref\CachazoAZ{
  F.~Cachazo, M.~Spradlin and A.~Volovich,
  ``Four-Loop Cusp Anomalous Dimension From Obstructions,''
  Phys.\ Rev.\  D {\bf 75}, 105011 (2007)
  [arXiv:hep-th/0612309].
}

\lref\ReyIK{
  S.~J.~Rey and J.~T.~Yee,
  ``Macroscopic strings as heavy quarks in large N gauge theory and  anti-de
  Sitter supergravity,''
  Eur.\ Phys.\ J.\  C {\bf 22}, 379 (2001)
  [arXiv:hep-th/9803001].
}

\lref\MaldacenaIM{
  J.~M.~Maldacena,
  ``Wilson loops in large N field theories,''
  Phys.\ Rev.\ Lett.\  {\bf 80}, 4859 (1998)
  [arXiv:hep-th/9803002].
}

\lref\BernAP{
  Z.~Bern, L.~J.~Dixon, D.~A.~Kosower, R.~Roiban, M.~Spradlin, C.~Vergu and A.~Volovich,
  ``The Two-Loop Six-Gluon MHV Amplitude in Maximally Supersymmetric Yang-Mills
  Theory,''
  arXiv:0803.1465 [hep-th].
}

\lref\Drufour{
  J.~M.~Drummond, J.~Henn, G.~P.~Korchemsky and E.~Sokatchev,
  ``The hexagon Wilson loop and the BDS ansatz for the six-gluon amplitude,''
  arXiv:0712.4138 [hep-th].
}

\lref\Drufive{
  J.~M.~Drummond, J.~Henn, G.~P.~Korchemsky and E.~Sokatchev,
  ``Hexagon Wilson loop = six-gluon MHV amplitude,''
  arXiv:0803.1466 [hep-th].
}

\lref\nastone{
  D.~Astefanesei, S.~Dobashi, K.~Ito and H.~Nastase,
  ``Comments on gluon 6-point scattering amplitudes in N=4 SYM at strong
  coupling,''
  JHEP {\bf 0712}, 077 (2007)
  [arXiv:0710.1684 [hep-th]].
}

\lref\Moroone{
  H.~Itoyama, A.~Mironov and A.~Morozov,
  ``Boundary Ring: a way to construct approximate NG solutions with polygon
  boundary conditions: I. $Z_n$-symmetric configurations,''
  arXiv:0712.0159 [hep-th].
}

\lref\Morotwo{
  H.~Itoyama and A.~Morozov,
  ``Boundary Ring or a Way to Construct Approximate NG Solutions with Polygon
  Boundary Conditions. II. Polygons which admit an inscribed circle,''
  arXiv:0712.2316 [hep-th].
}

\lref\KruczenskiCY{
  M.~Kruczenski, R.~Roiban, A.~Tirziu and A.~A.~Tseytlin,
  ``Strong-coupling expansion of cusp anomaly and gluon amplitudes from quantum
  open strings in $AdS_5 x S^5$''
  Nucl.\ Phys.\  B {\bf 791}, 93 (2008)
  [arXiv:0707.4254 [hep-th]].
}

\lref\KomargodskiER{
  Z.~Komargodski and S.~S.~Razamat,
  ``Planar quark scattering at strong coupling and universality,''
  JHEP {\bf 0801}, 044 (2008)
  [arXiv:0707.4367 [hep-th]].
}

\lref\McGreevyKT{
  J.~McGreevy and A.~Sever,
  ``Quark scattering amplitudes at strong coupling,''
  JHEP {\bf 0802}, 015 (2008)
  [arXiv:0710.0393 [hep-th]].
}

\lref\ItoZY{
  K.~Ito, H.~Nastase and K.~Iwasaki,
  ``Gluon scattering in ${\cal N}=4$ Super Yang-Mills at finite temperature,''
  arXiv:0711.3532 [hep-th].
}

\lref\OzQR{
  Y.~Oz, S.~Theisen and S.~Yankielowicz,
  ``Gluon Scattering in Deformed N=4 SYM,''
  arXiv:0712.3491 [hep-th].
}

\lref\KruczenskiFB{
  M.~Kruczenski,
  ``A note on twist two operators in N = 4 SYM and Wilson loops in Minkowski
  signature,''
  JHEP {\bf 0212}, 024 (2002)
  [arXiv:hep-th/0210115].
}

\lref\DixonTU{
  L.~J.~Dixon,
  ``Gluon scattering in N=4 super-Yang-Mills theory from weak to strong
  coupling,''
  arXiv:0803.2475 [hep-th].
}

\lref\StermanQN{
  G.~Sterman and M.~E.~Tejeda-Yeomans,
  ``Multi-loop amplitudes and resummation,''
  Phys.\ Lett.\  B {\bf 552}, 48 (2003)
  [arXiv:hep-ph/0210130].
}

\lref\JevickiPK{
  A.~Jevicki, C.~Kalousios, M.~Spradlin and A.~Volovich,
  ``Dressing the Giant Gluon,''
  JHEP {\bf 0712}, 047 (2007)
  [arXiv:0708.0818 [hep-th]].
}

\Title{\vbox{\baselineskip12pt \hbox{SPIN-08/18} \hbox{
ITP-UU-08/20} }} {\vbox{\centerline{ Lectures on Scattering
Amplitudes via $AdS/CFT$ }}}
\bigskip
\centerline{ Luis F. Alday}

\bigskip
\centerline{\it
Institute for Theoretical Physics and Spinoza
Institute} \centerline{Utrecht University, 3508 TD Utrecht, The
Netherlands}

\vskip .3in \noindent
We review recent progress on computing scattering
amplitudes of planar ${\cal N}=4$ super Yang-Mills at strong coupling by using
the AdS/CFT duality. We do explicit computations by using both, dimensional regularization and a cut-off in the radial direction.  Up to an additive constant independent on the kinematics, the finite piece of the amplitude is the same in both regularizations. The later scheme is particularly appropriate for understanding the conformal properties of the amplitudes.\foot{Based on lectures given by the author at several places.}


 \Date{ }


\newsec{Introduction}

In these notes we study  gluon scattering amplitudes of planar
maximally super-symmetric, ${\cal N}=4$, Yang-Mills (MSYM). In
this theory one can go much further in the computation of
scattering amplitudes than in other gauge theories, such as QCD, but at the same time we expect
that these amplitudes can can teach us something about QCD amplitudes. On one
hand, perturbative computations are simpler. In the last few years
there have been an enormous progress in the computation of MSYM
scattering amplitudes. On the other hand, the strong coupling
limit of the theory can be studied by means of the $AdS/CFT$
duality by considering a dual weakly coupled string sigma model.

The main aim of these notes is to report on recent progress
 in the application of the  $AdS/CFT$
duality to compute gluon scattering amplitudes of planar MSYM at
strong coupling \amf\ \ams\ .

These notes are organized as follows. In the next section we
briefly describe some weak coupling perturbative results for the
  scattering amplitudes in planar MSYM. In section three we
explain how the $AdS/CFT$ duality can be used   to compute
scattering amplitudes at strong coupling. The amplitudes are
IR divergent, so a regulator needs to be introduced in order to
define them properly. We start by introducing a D-brane as
infra-red (IR) regulator in order to set-up our calculation.
Actual computations, however, are done in the super-gravity analog
of dimensional regularization, since it is easier to proceed in
this scheme and besides, we want to compare our results with the
perturbative results which were computed using dimensional
regularization. We show in some detail how our prescription works
for the scattering of four gluons and repeat the same computation
by using a cut-off in the radial direction as IR regulator. At the
end of section three we emphasize the role of a dual conformal
symmetry and present a simple argument leading to a (dual)
conformal Ward identity which fixes the form of the amplitude for
the scattering of four and five gluons. In section four we review
a recent conjectured duality between scattering amplitudes and
Wilson loops and we use the proved one loop duality in order to
test a particular guess for the form of the scattering amplitudes.
Finally we end up with some conclusions and a list of open
problems and future directions.

\newsec{Perturbative planar MSYM scattering amplitudes}

In this section we briefly discuss the progress during the last
few years in computing perturbative planar scattering amplitudes
of MSYM. \foot{Among the vast literature on the subject we refer
the reader to \BernZX\bds\AnastasiouKJ\DixonTU  and references
therein for a detailed account of the material exposed in these
notes.}

Gluon states $|{\cal G} \rangle=|h,p^\mu,a \rangle$ are
characterized by their helicity $h=\pm 1$, four momenta $p^\mu$
and color indices $a$ in the adjoint representation. Generic
amplitudes  depend in a complicated manner (even at tree level!)
on these.

In the limit of a large number of colors it is useful to write the
amplitudes into a color decomposed form. Denoting by ${\cal
A}^{(L)}_n$ the $L-$loop, $n-$point amplitude we have

\eqn\colordec{ {\cal A}^{(L)}_n \approx g^{n-2}(g^2 N)^L \sum_\rho
\Tr(T^{a_{\rho(1)}}...T^{a_{\rho(n)}})
A_n^{(L)}(\rho(1),...,\rho(n))
}
where the sum runs over non cyclic
permutations, $N$ denotes the number of colors and $g$ the gauge
theory coupling constant. This decomposition clearly separates the
color structure from the kinematics. The leading color ordered
amplitude $A_n^{(L)}$ will hence depend only on the momenta and
the helicities of the particles undergoing the scattering.

On shell amplitudes of massless gauge theories in four dimensions,
such as $MSYM$, contain IR singularities.\foot{These amplitudes
can be used as building blocks for constructing well defined, IR
finite, physical observables.} Hence, in order to define them, one
needs to introduce a regulator. Commonly, a version of dimensional
regularization is used, {\it i. e.} we consider the theory in
$D=4-2\epsilon$ dimensions. For instance, the one loop scattering
amplitude for four gluons contains a factor of the form

\eqn\oneloopfourgluons{I_4^{(1)}=\int {d^Dp \over
p^2(p-k_1)^2(p-k_1-k_2)^2(p+k_4)^2}}

We recognize two kind of divergences. First, from the region
$p^\mu \sim 0$. These are called soft divergences, since they are
due to the interchange of soft (with very low momentum) gluons
between external gluons. The second class comes from the region
$p^\mu \sim \alpha k^\mu_i$ and are called collinear divergences,
since in this case the momentum of the gluon interchanged is
proportional to the momentum of one of the external gluons.  We
can also have soft and collinear divergencies.

Once a regulator is introduced the amplitudes are finite but will
depend explicitly on such regulator. IR divergences manifest as
poles in $\epsilon$, {\it e.g.} $A_n^{(L)}\simeq {1 \over
\epsilon^{2L}}+...$.

As already mentioned, the color ordered amplitudes $A_n^{(L)}$
depend also on the helicities of the external gluons. In MSYM,
scattering amplitudes satisfy super-symmetric Ward identities that
imply the vanishing of the amplitudes for particular choices of
the gluon helicities. For instance, it can be shown that the
amplitude vanishes when all helicities, or all but one, are plus

\eqn\susyward{{\cal A}(+++...+)={\cal A}(-++...+)=0}

The first non trivial example is the one with two minuses and the
rest plus, ${\cal A}(--++...+)$. Such amplitudes are called
maximally helicity violating (MHV) amplitudes.

The simplicity of studying MHV amplitudes is given by the fact
that they contain a single Lorentz structure, which is captured by
the tree level amplitude, to all orders in perturbation theory. It
is then convenient to factor out the tree level amplitude and
study the reduced amplitude

\eqn\reducedamp{M_n^{(L)}(\epsilon)={A_n^{(L)}(\epsilon) \over
A_n^{(0)}}}
which depends only on the kinematical invariants and the
regulator.

Two loops computations show that at this order, the amplitudes can
be written in terms of lower order amplitudes, for instance

\eqn\recrel{M_4^{(2)}(\epsilon)={1 \over 2}
\left(M_4^{(1)}(\epsilon)
\right)^2+f^{(2)}(\epsilon)M_4^{(1)}(2\epsilon)+C^{(2)}+{\cal
O}(\epsilon)}
such relation is non trivial, since the constants
$f^{(2)}(\epsilon)$ and $C^{(2)}$ don't depend on the kinematics.
Notice that in order to check such relation, $M_4^{(1)}(\epsilon)$
should be computed up to $\epsilon^2$ order. An analogous
relation, with the same $f^{(2)}(\epsilon)$ and $C^{(2)}$, is
satisfied by the five point scattering amplitude.

Based on explicit computations, plus the well known structure of divergences and consistency checks when taking collinear limits, Bern, Dixon and Smirnov  (BDS) \bds ~ proposed the
following ansatz for the MHV scattering amplitudes of any number
of gluons at any loop order
\eqn\BDSansatz{
{\cal M}_n \equiv
1+\sum_{\ell=1} \alpha^\ell M_n^{(\ell)}=\exp \left[ \sum_{\ell=1}
\alpha^\ell \left( f^{(\ell)}(\epsilon) M_4^{(1)}(\ell \epsilon)
+C^{(\ell)} +{\cal O}(\epsilon) \right)\right]
}
where
\eqn\bdsconst{f^{(\ell)}(\epsilon)=f^{(\ell)}_0+ \epsilon
f^{(\ell)}_1+\epsilon^2 f^{(\ell)}_2,~~~~~~~~~~~~\alpha \approx
\lambda \mu^{2\epsilon}
}
$\alpha$ is proportional to the 't Hooft coupling constant and
keeps track of the perturbation order. The IR regulator scale
$\mu$ accounts for the fact that in dimensions different from four
the coupling constant is not dimensionless.

The structure of the IR divergent terms in \BDSansatz\ was determined in
\StermanQN . The non-trivial conjecture is that the finite pieces
are given by the same functions that characterize the IR divergent terms.
 Hence, the BDS ansatz give us the (log of the)
amplitude at any loop order and for any number of gluons in terms
of the one loop amplitude, up to a set of numbers that have to be
computed (for instance by computing amplitudes explicitly for a
low number of points). We will see that there are good reasons for thinking that
BDS is correct for four and five gluons but that starting at six gluons it
does not give us the right answer.






We will be particularly interested in the scattering of four
gluons, in that case the BDS ansatz reduces to the simple expression
\eqn\bdsfour{\eqalign{  A_4 =A_{tree}\left(A_{div,s} \right)^2
\left(A_{div,t} \right)^2 \exp \left( {f(\lambda) \over
8}\left(\log {s \over t}\right)^2+const \right) \cr
A_{div,s}=\exp \left( -{1 \over 8 \epsilon^2}
f^{(-2)}\left({\lambda \mu^{2\epsilon} \over s^\epsilon} \right) -
{1 \over 4 \epsilon} g^{(-1)}\left({\lambda \mu^{2\epsilon} \over
s^\epsilon } \right) \right)}}
where $s$ and $t$ are the usual Mandelstan variables for the
scattering of four particles. The amplitude has two pieces, a
divergent one and a finite one. The leading divergent piece is
characterized by the so called cusp anomalous dimension
$f(\lambda) =(\lambda \partial_\lambda)^2 f^{(-2)}(\lambda)$,
while the subleading divergent part is controlled by the function
$g(\lambda)$, sometimes called collinear anomalous dimension.
Notice that the cusp anomalous dimension also controls the
kinematical dependent factor of the finite piece, proportional to
$(\log {s \over t})^2$. Much is known about $f(\lambda)$, in
particular, by independent means, its strong coupling behavior has
been computed \gkp\

\eqn\fstrong{f(\lambda) = {\sqrt{\lambda}\over
\pi}+...,~~~~~~~~~\lambda \gg 1}

In the next section we explain how to compute gluon
scattering amplitudes of planar MSYM at strong coupling by using
the $AdS/CFT$ duality.  We will show that the four gluons  answer
is indeed given by \bdsfour\ at strong coupling.

\newsec{Gluon scattering amplitudes at strong coupling}

In order to attack the problem of computing scattering amplitudes
at strong coupling we will make use of the $AdS/CFT$ duality \MaldacenaRE . This
duality, expresses the equivalence between four dimensional MSYM
and type IIB string theory on $AdS_5 \times S^5$. The duality
provides us with a dictionary between the parameters on both sides

\eqn\adsdict{\sqrt{\lambda} \equiv \sqrt{g^2_{YM} N}={ R^2 \over
\alpha' },~~~~~~~~~~~~{1 \over N} \sim g_s}
relating in this way the 't Hooft coupling constant to the radius
of compactification of the $S^5$ and $AdS_5$ in string units, and
the inverse of the number of colors to the string coupling
constant.

Thus, we see that in the limit of a large number of colors,
strings don't split or join, and we can describe string theory by
a non linear sigma model. In the regime in which $\lambda$ is very
large, this sigma model is weakly coupled.

As in the gauge theory, we need to introduce a regulator to
properly define scattering amplitudes. In order to set-up our
computation we introduce a D-brane as IR regulator.

\subsec{Set-up of the computation}

 As a first IR reglulator we consider a D-brane
localized in the radial direction.
In other words, we start with the $AdS_5$ metric written in Poincare coordinates
\eqn\poinc{ds^2 = R^2 {dx_{3+1}^2 +dz^2 \over z^2}}
and we place a D-brane at some fixed large value of $z=z_{IR}  $ and
extending along the $x_{3+1}$ coordinates. The asymptotic states are
open strings that end on the D-brane. We then consider the scattering
of these open strings.

The proper momentum of the strings is $k_{pr}=k z_{IR}/R$, where
$k$ is the momentum conjugate to $x_{3+1}$, plays the role of
gauge theory momentum and will be kept fixed as we take away the
IR cut-off, $z_{IR} \to \infty$. Therefore, due to the warping of the metric, the
proper momentum is very large, so we are considering the
scattering of strings at fixed angle with very large momentum.
Amplitudes in such regime were studied in flat space by Gross,
Mende and Manes, \refs{\GrossKZA,\grossmanes}.
The key feature of their computation is that the amplitudes
are dominated by a saddle point of the classical action. In our
case we need to consider classical strings on $AdS$.

We need then to consider a world-sheet with the topology of a disk
with vertex operator insertions on its boundary, which correspond
to the external states (see fig. 1). Each color ordered amplitude corresponds
to a disk amplitude with a fixed ordering of the open string vertex operators.

\ifig\branesc{World-sheet corresponding to the scattering of four
open strings.} {\epsfxsize3in\epsfbox{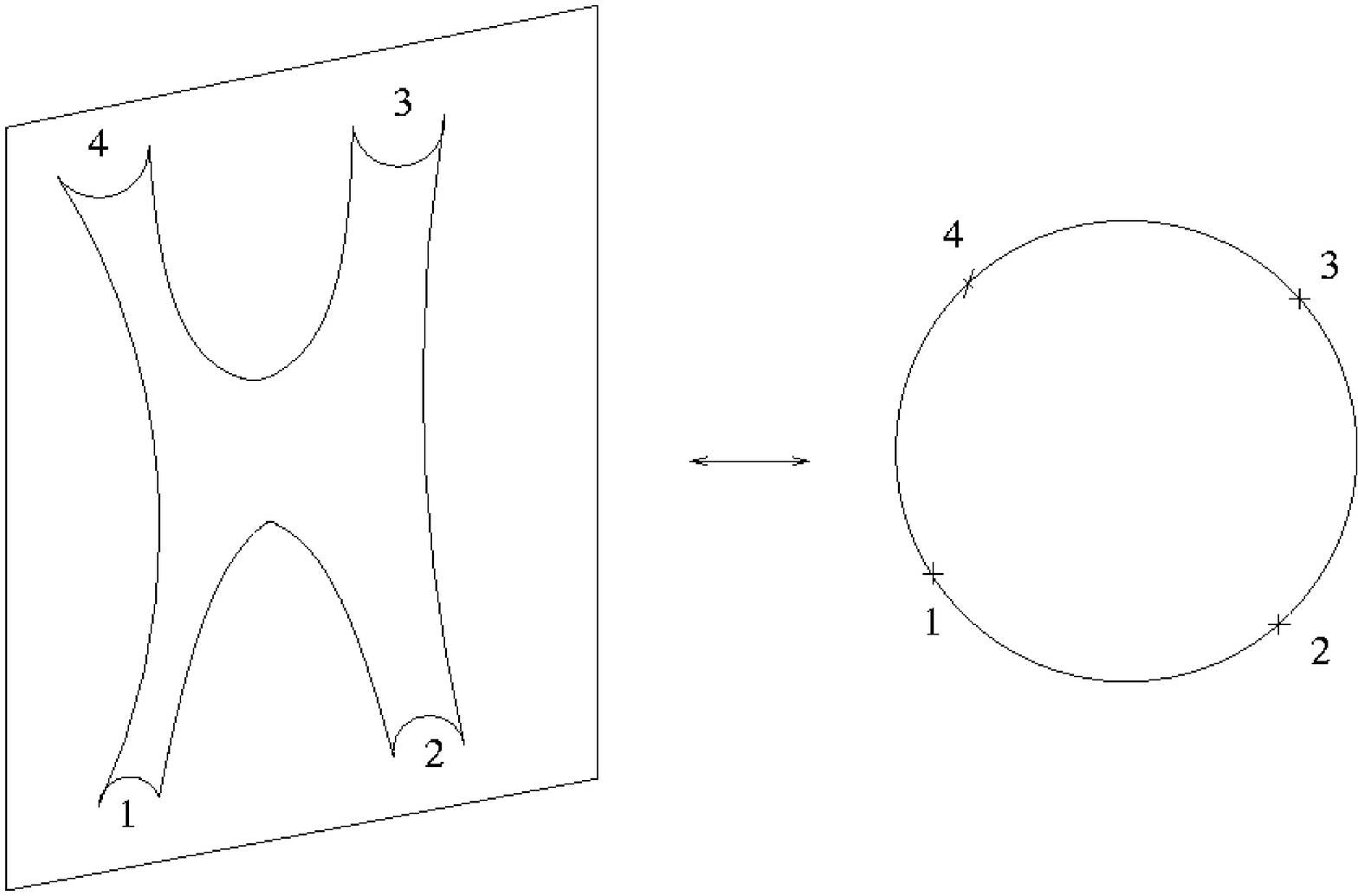}}

The boundary conditions for the world-sheet are the following: In
the vicinity of a vertex operator, the momentum of the external
state fixes the form of the solution and, as the open strings are
attached to the D-brane, $z=z_{IR}$ at the boundary.

In order to state more simply the boundary conditions for the
world-sheet, it is convenient to describe the solution in terms of
T-dual coordinates $y^\mu$, defined as

\eqn\tdual{ds^2=w^2(z)dx_\mu dx^\mu ~~~\rightarrow ~~~
\partial_\alpha y^\mu = i w^2(z) \epsilon_{\alpha \beta} \partial_\beta x^\mu}
Note that we don't T-dualize the radial direction $z$. The
boundary conditions for the original coordinates $x^\mu$, which are
that they carry momentum $k^\mu$, translates into the condition
that $y^\mu$ has "winding"

\eqn\ybc{\Delta y^\mu=2 \pi k^\mu}
After defining $r=R^2/z$ we end up again with the $AdS_5$ metric

\eqn\dualads{ds^2=R^2 {dy_\mu dy^\mu+dr^2 \over r^2}}
Now the boundary of the world-sheet is located at $r=R^2/z_{IR}$
and is a particular line constructed as follows (see fig. 2)

\ifig\sixlines{Polygon of light-like segments corresponding to the
momenta of the external particles.}
{\epsfxsize1.5in\epsfbox{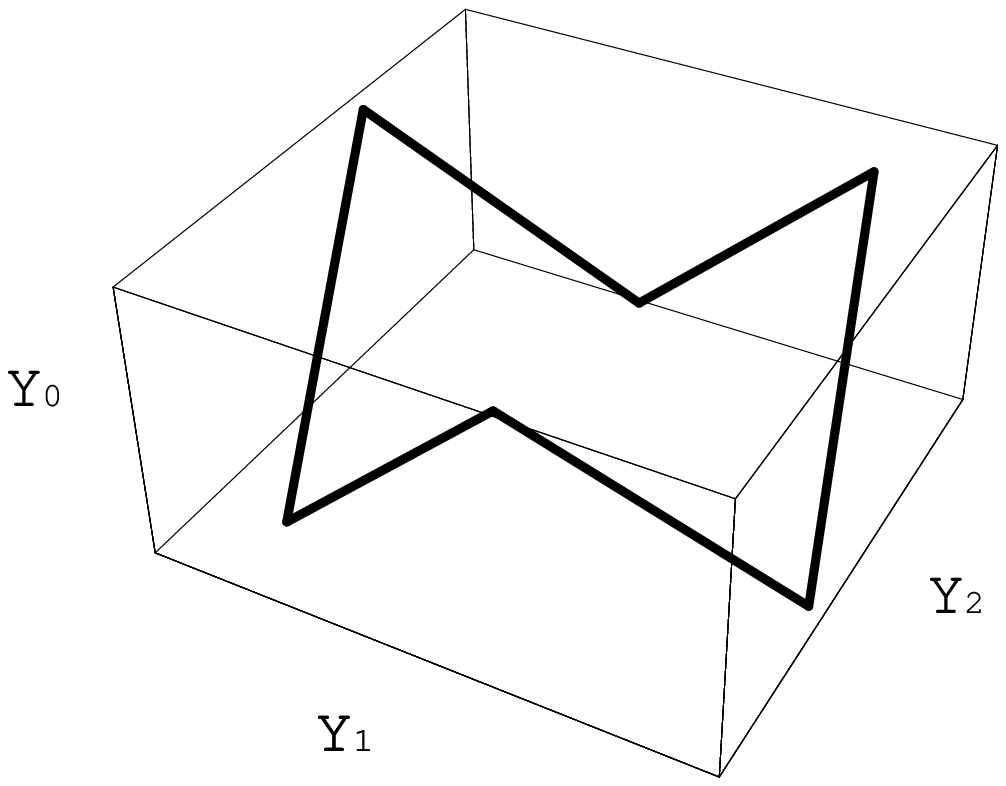}}

\item{$\bullet$} For each particle of momentum $k^\mu$, draw a
segment joining two points separated by $\Delta y^\mu=2 \pi
k^\mu$.

\item{$\bullet$} Concatenate the segments according to the
insertions on the disk (corresponding to a particular color
ordering)

\item{$\bullet$} As gluons are massless, the segments will be
light-like. Due to momentum conservation, the diagram is closed.

The world-sheet, when expressed in T-dual coordinates, will then
end in such sequence of light-like segments (see fig. 3) located
at $r=R^2/z_{IR}$

\ifig\orvsdual{Comparison of the world sheet in original and
T-dual coordinates.} {\epsfxsize3in\epsfbox{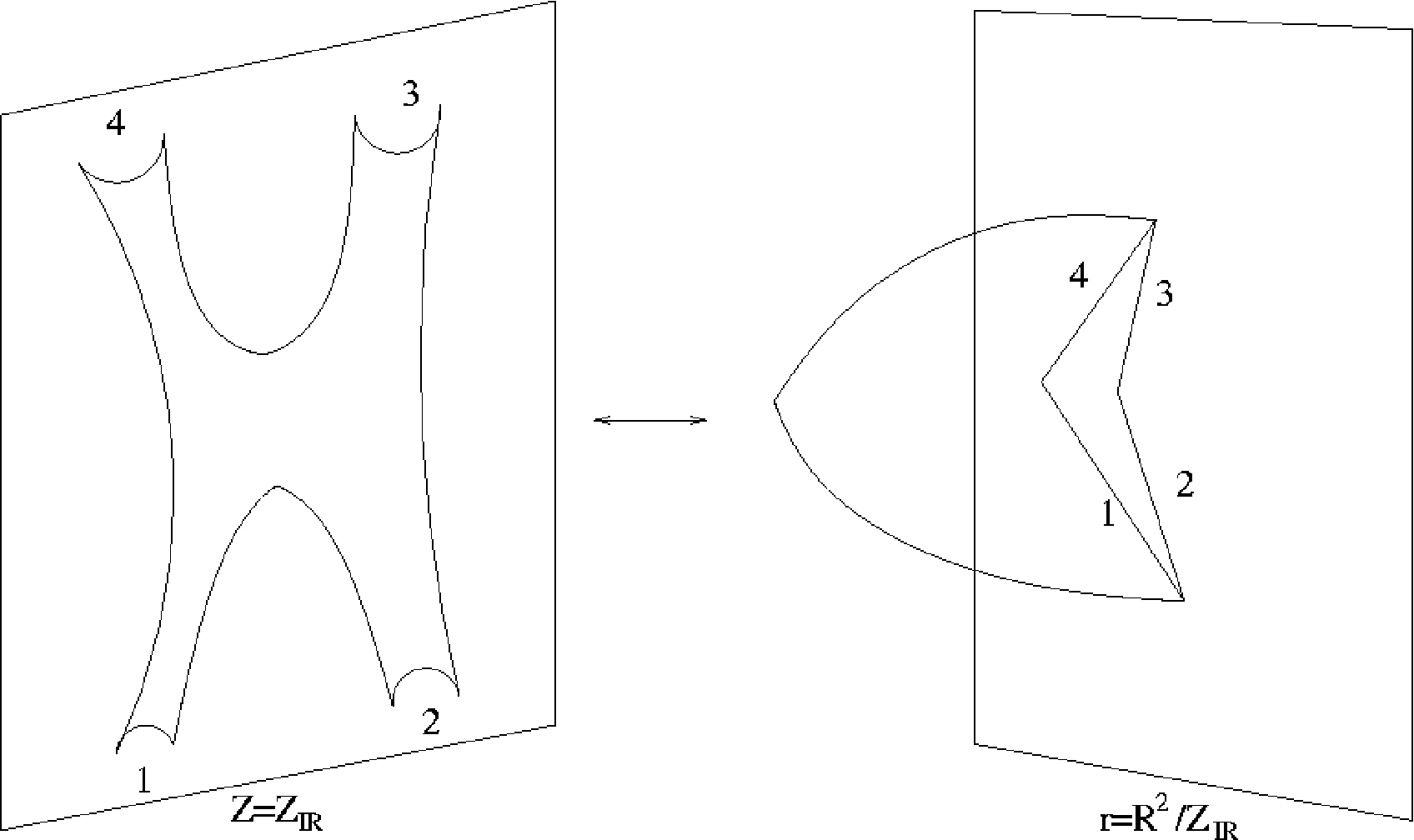}}

As we take away the IR cut-off, $z_{IR} \rightarrow \infty$, the
boundary of the world-sheet moves towards the boundary of the T-dual
metric, at $r=0$. At leading order in the strong coupling expansion,
the computation that we are doing is formally
the same as the one we would do if we were computing
the expectation value of a Wilson loop given by a
sequence of light-like segments.

Our prescription is then that the leading exponential behavior of
the $n-$point scattering amplitude is given by the area $A$ of the
minimal surface that ends on a sequence of light-like segments on
the boundary
\eqn\finalpresc{{\cal A}_n \sim e^{-{\sqrt{\lambda} \over 2
\pi}A(k_1,...,k_n)}}
The area $A$ contains the kinematical information through its
boundary conditions.

We stress that our computation is blind to the polarization of the
gluons, which contribute to prefactors in \finalpresc\ and are
subleading in $1/\sqrt{\lambda}$.

In the following, we will show in detail how our prescription
works for the scattering of four gluons and compare our results
with field theory expectations.

\subsec{$n=4$ case}

Consider the scattering of two particles into two particles,
$k_1+k_3 \rightarrow k_2+k_4$ and define the usual Mandelstam
variables

\eqn\mandel{s=-(k_1+k_2)^2,~~~~~t=-(k_2+k_3)^2}
According to our prescription we need to find the minimal surface ending in the following light-like polygon

\ifig\square{Polygon corresponding to the scattering of four
gluons} {\epsfxsize1.5in\epsfbox{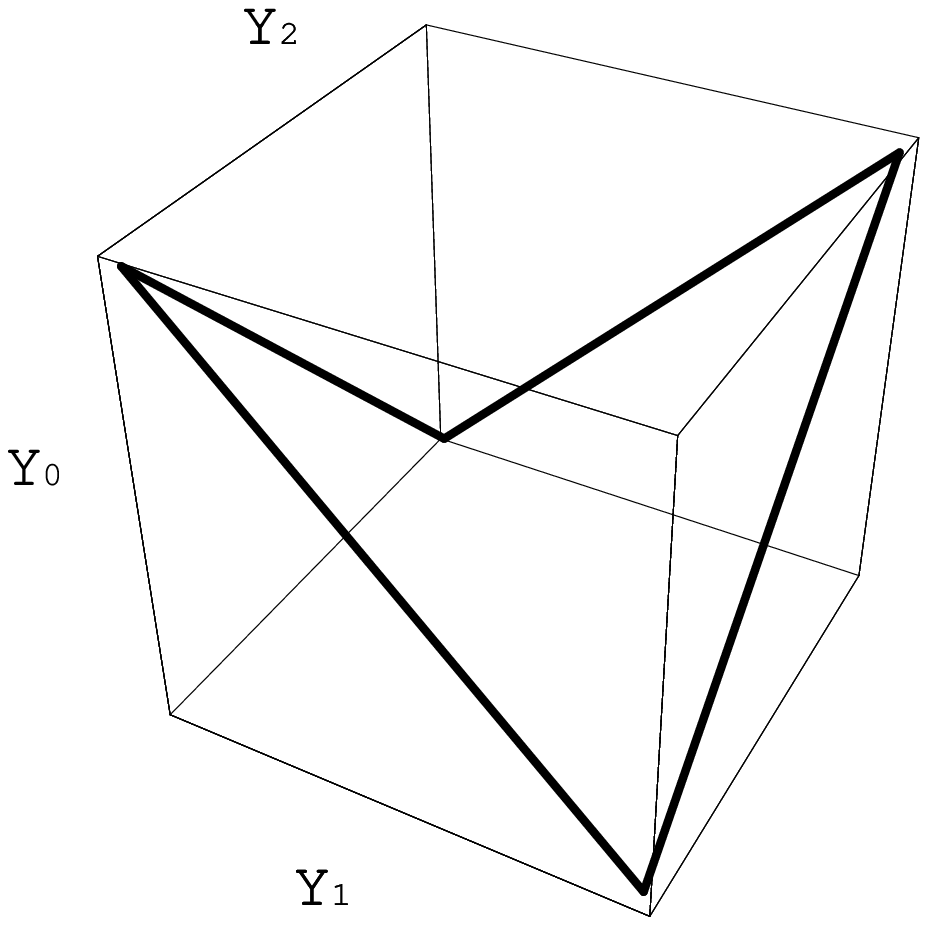}}

In order to write the Nambu-Goto action it is convenient to use Poincare coordinates $(r,y_0,y_1,y_2)$, setting $y_3=0$ and parametrize the surface by its projection to the $(y_1,y_2)$ plane. In this case we obtain an action for two fields, $r$ and $t$, living in the space parametrized by $y_1$ and $y_2$

\eqn\ngedge{ S = {R^2 \over2 \pi
} \int dy_1 dy_2 { \sqrt{ 1 + ( \partial_i r)^2 - (
\partial_i y_0)^2 - (
\partial_1 r \partial_2y_0 - \partial_2 r \partial_1 y_0 )^2 } \over
r^{ 2 } } }
The classical equations of motion should then be supplemented by the appropriate boundary conditions. We consider
first the case with $s=t$ where the projection of the Wilson lines
 is a square. By scale invariance, we
can change the size of the square.
We choose the edges of the square to be at $y_1,y_2=\pm 1$. The
boundary conditions are then given by
\eqn\squarebc{ r(\pm 1,y_2)=r(y_1,\pm 1)=0,~~~~y_0(\pm 1,y_2)=\pm
y_2,~~~y_0(y_1,\pm 1)=\pm y_1}
The form of the solution near each of the cusps can be obtained
from the single cusp solution of \KruczenskiFB . Making educated
guesses satisfying the boundary conditions and with the correct
properties near the cusps we propose

\eqn\squaresol{y_0(y_1,y_2)=y_1
y_2,~~~~~r(y_1,y_2)=\sqrt{(1-y_1^2)(1-y_2^2)} } Remarkably it
turns out to be a solution of the equations of motion. However, in
order to capture the kinematical dependence of \bdsfour we need to
consider more general solutions with $s \neq t$. In this case the
projection of the surface to the $(y_1,y_2)$ plane will not be an
square but a rombus, with $s$ and $t$ given by the square of the
distance between opposite vertices, as shown in fig. 5.

\ifig\squarevsrombus{Projection to the plane $(y_1,y_2)$ of the
surface for the cases $s=t$ and $s \neq t$.}
{\epsfxsize2.5in\epsfbox{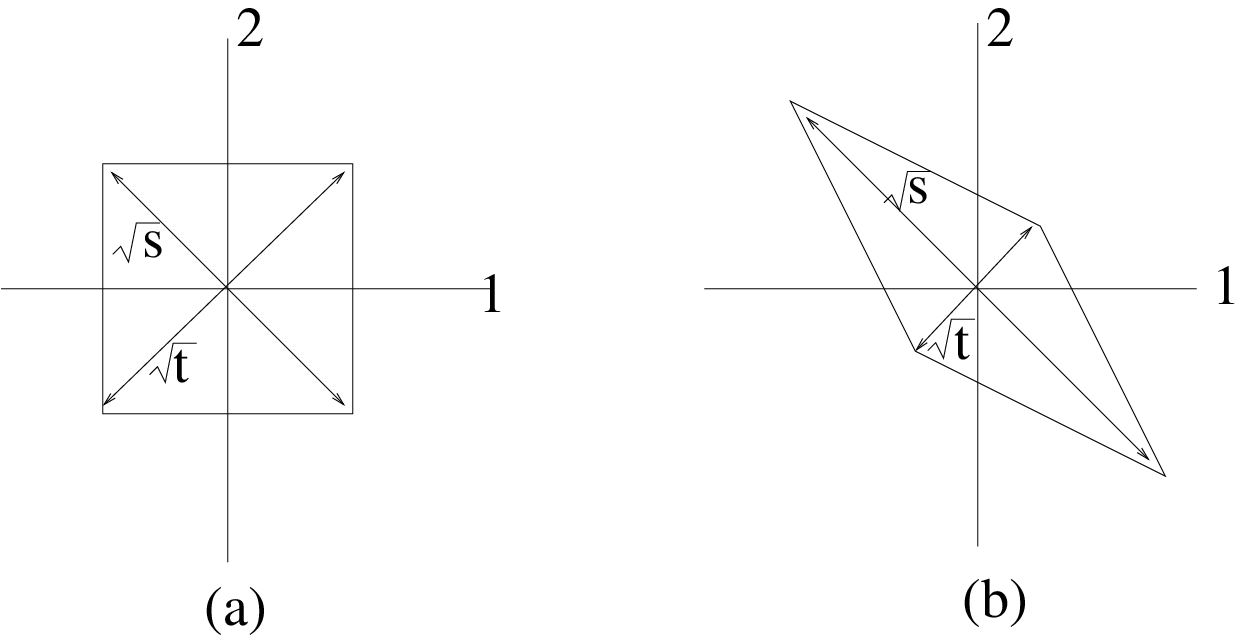}}

In order to find the solution for this more general case, it is
instructive to study the surface \squaresol\ in terms of embedding
coordinates. These are coordinates where we view $AdS_5$ as the
following surface embedded in $R^{2,4}$

\eqn\embcoordl{-Y_{-1}^2-Y_{0}^2+Y_1^2+Y_2^2+Y_3^2+Y_4=-1}
The relation between these and Poincare coordinates  is
\eqn\intermofpu{\eqalign{ Y^\mu & = { y^\mu \over r} ~,~~\mu
=0,\cdots ,3 ~ \cr Y_{-1} + Y_4 & = { 1 \over r } ~,~~~~~~~~Y_{-1}
- Y_4 = { r^2 + y_\mu y^\mu \over r } }}
The surface \squaresol\  is then given by

\eqn\squaresol{Y_0 Y_{-1}=Y_1 Y_2,~~~~~~Y_3=Y_4=0}
Once we have written our solution in embedding coordinates, we notice that we can apply $SO(2,4)$ transformations, that are linearly realized in this coordinates, in order to obtain new solutions.
This $SO(2,4)$  symmetry is sometimes referred to
as "dual conformal symmetry" and should not be confused with the
 original $SO(2,4)$ symmetry associated to the original $AdS$ space. It was first
  observed in
 computations at weak coupling in \dualconf .

Solutions with $s \neq t$ can be obtained by starting with \squaresol\ and performing a boost in the $0-4$ direnction

\eqn\squaresol{Y_0 Y_{-1}=Y_1 Y_2,~~~Y_4=0~~~\rightarrow~~~Y_4-v
Y_0=0,~~~\sqrt{1-v^2}Y_0 Y_{-1}=Y_1,Y_2}
Hence, we end up with a two parameters solutions, one related to the size
of the initial square and another related to the boost parameter.

\eqn\soluvrom{\eqalign{r={a \over \cosh u_1 \cosh
u_2+b \sinh u_1 \sinh u_2},~~~~ y_0= {a \sqrt{1+b^2} \sinh u_1 \sinh
u_2 \over \cosh u_1 \cosh u_2+b \sinh u_1 \sinh u_2} \cr y_1={a
\sinh u_1 \cosh u_2 \over \cosh u_1 \cosh u_2+b \sinh u_1 \sinh
u_2},~~~~ y_2={a \cosh u_1 \sinh u_2 \over \cosh u_1 \cosh u_2+b
\sinh u_1 \sinh u_2} }}
where we have written the surface as a solution to the equations
of motion in conformal gauge

\eqn\confact{iS=-{R^2 \over 2 \pi } \int {\cal L} =-{R^2 \over 2 \pi }
\int du_1 du_2  { 1 \over 2 }
 { \left( \partial r\partial r +  \partial y_\mu \partial y^\mu
 \right)\over r^2 }}
$a$ and $b$ encode the kinematical information of the scattering as follows

\eqn\sandt{-s(2 \pi)^2 = {8 a^2 \over (1-b)^2},~~~~~-t (2 \pi)^2 ={8 a^2 \over (1+b)^2},
 ~~~~~{ s \over t } = { (1+b)^2 \over
(1-b)^2 } }
According to the prescription, we should now plug the classical solution into the classical action in order to obtain the four point scattering amplitude at strong coupling. However, in doing so, we obtain a divergent answer. That is of course the case, since we have taken the IR regulator away, in order to obtain a finite answer we need to reintroduce a regulator.

\bigskip

\noindent i.-Dimensional regularization

\bigskip

Gauge theory amplitudes are regularized by considering the theory in $D=4-2\epsilon$ dimensions. More precisely \BernAQ, one starts with ${\cal N}=1$ in ten dimensions and then dimensionally reduce to $4-2\epsilon$ dimensions. For integer $2\epsilon$ this is precisely the low energy theory living on a $Dp-$brane, where $p=3-2\epsilon$. We regularize the amplitudes at strong coupling by considering the gravity dual of these theories. The string frame metric is
 \eqn\gravdime{\eqalign{ds^2=f^{-1/2}dx_{4-2\epsilon}^2+f^{1/2}\left[ dr^2+r^2 d\Omega^2_{5+2\epsilon} \right],~~~~~f=(4 \pi^2 e^\gamma)^\epsilon \Gamma(2+\epsilon) \mu^{2\epsilon} \frac{\lambda}{r^{4+2\epsilon}}}}
We are then led to the following action
\eqn\actioncorr{S=\frac{\sqrt{c_\epsilon \lambda} \mu^\epsilon }{2\pi} \int \frac{{\cal L}_{\epsilon=0}}{r^\epsilon}}
Where ${\cal L}_{\epsilon=0}$ is the Lagrangian density for
$AdS_5$. The presence of the factor $r^\epsilon$ will have two
important effects. On one hand, previously divergent integrals
will now converge. On the other hand, the equations of motion will
now depend on $\epsilon$ and we were not able to compute the full
solution for arbitrary $\epsilon$. However, we are interested in
computing the amplitude only up to finite terms as we take
$\epsilon \rightarrow 0$. In that case, it turns out to be
sufficient to plug the original solution into the
$\epsilon$-deformed action \foot{Up to a contribution from the
regions close to the cusps that add an unimportant additional
constant term.}. After performing the integrals we obtain:

\eqn\actionint{S \approx \sqrt{\lambda} \frac{\mu^\epsilon}{a^\epsilon} ~ _2 F_1\left( \frac{1}{2},-\frac{\epsilon}{2},\frac{1-\epsilon}{2};b^2 \right)}
Expanding in powers of $\epsilon$ we get the final answer

\eqn\finalfour{\eqalign{{\cal A}=e^{i S}=\exp \left[i S_{div}+\frac{\sqrt{\lambda}}{8\pi}\left(\log{\frac{s}{t}} \right)^2+\tilde{C} \right] \cr
S_{div}=2S_{div,s}+2S_{div,t}\cr
i S_{div,s}=-\frac{1}{\epsilon^2}\frac{1}{2\pi}\sqrt{\frac{\lambda \mu^{2\epsilon}}{(-s)^\epsilon}} -\frac{1}{\epsilon}\frac{1}{4\pi}(1-\log 2) \sqrt{\frac{\lambda \mu^{2\epsilon}}{(-s)^\epsilon}}  }}
This should be compared with the field theory expectations
\bdsfour\ .  We notice that the general structure is in perfect
agreement with the BDS ansatz. Once we use the strong coupling
behavior for the cusp anomalous dimension \fstrong\ we see that
the leading divergence has the correct coefficient, besides, from
\finalfour\ one could extract the strong coupling behavior of the
function $g(\lambda)$. Finally,  the
kinematical part of the finite piece agrees exactly with the four gluons BDS
prediction.

\bigskip

\noindent ii.-Radial cut-off

\bigskip

A more common regularization scheme for computing minimal areas in
$AdS$ is to introduce a cut-off in the radial direction. The
correct procedure is to impose the boundary conditions at some
small $r=r_c$. It turns out, however, that in order to compute the
finite piece as $r_c \rightarrow 0$ it suffices to use the
original solution and cut the integral giving the area at
$r=r_{c}$\foot{The situation is completely analogous that what
happened when using dimensional regularization.}

In order to compute the regularized area for the scattering of
four gluons it is convenient to work in conformal gauge. In this
case, the problem boils down to compute the area enclosed by the
curve

\eqn\rcut{\frac{a}{\cosh u_1 \cosh u_2+b \sinh u_1 \sinh
u_2}=r_{c}}
The resulting integrals are pretty tedious. One way to proceed is
by expanding the integrand in power series of $r_c / a$ and
integrating term by term. Equivalently, one can use Green's
theorem and express the area as a one dimensional integral over
the boundary of the world-sheet. Finally, applying various known
identities between $ArcSech$ and logarithms we arrive to the final
expression for the area

\eqn\nfourcutoff{\eqalign{i S
=-\frac{\sqrt{\lambda}}{2\pi}A,~~~~~A=\frac{1}{4}\left(\log\left(\frac{r_c^2}{-8
\pi^2 s }\right)
\right)^2+\frac{1}{4}\left(\log\left(\frac{r_c^2}{-8 \pi^2 t
}\right) \right)^2-\frac{1}{4}\log^2(\frac{s}{t})+const. } }
Several comments are in order. First, notice that the structure of
infrared divergences is of the form $\log^2(r^2_c/s)$, and the
coefficient in front of double logs and the finite piece are the
same, and can be identified with the cusp anomalous dimension, as in
the case of dimensional regularization. Second, single logs are
absent. Finally, the finite piece agrees, up to an additive
constant, with the one obtained by using dimensional
regularization. Hence, the computation of amplitudes at strong
coupling does not need to be done by using dimensional
regularization.

The IR structure of amplitudes at strong coupling in the general
case of $n-$point amplitudes can easily be understood. Given the
cusp formed by a pair of neighboring gluons with momenta $k_i$
and $k_{i+1}$ we associate the kinematical invariant
$s_i=(k_i+k_{i+1})^2$. We expect the following structure for the
IR divergent part of the action

\eqn\sdiv{i S_{div}=-\frac{\sqrt{\lambda}}{2\pi}\sum_i
I(\frac{r_c^2}{s_{i,i+1}})}
where $I(\frac{r_c^2}{s_{i,i+1}})$ can be computed following
\BuchbinderHM\ but using a radial cut-off instead of dimensional
regularization.

\eqn\sdivterm{4 I = \int^1_\xi \int^1_\frac{\xi}{X^-} \frac{1}{X^-
X^+}=\frac{1}{2}\log^2 \xi,~~~~~~~\xi=\frac{r_c^2}{-8 \pi^2
s_{i,i+1}}}
Hence, when using a radial cut-off as regulator, we expect the
following structure for scattering amplitudes at strong coupling

\eqn\sfull{i S_n=-\frac{\sqrt{\lambda}}{16\pi}\sum_{i=1}^n
\log^2\left(\frac{r_c^2}{-8 \pi^2 s_{i,i+1}} \right)+Fin(k_i) }
It is easy to check that the general form of the amplitude for the
case $n=4$ is consistent with this general expression.

For the discussion below, it will be important to consider a radial cut-off that depends on the point at the boundary we are approaching, {\it i.e.} $r_c(x)$. In that case, the structure of the amplitude turns out to be as follows

\eqn\sfullge{i S_n=-\frac{\sqrt{\lambda}}{16\pi}\sum_{i=1}^n
\log^2\left(\frac{r_c^2(x_i)}{-8 \pi^2 s_{i,i+1}} \right)+Fin(k_i)+\sum_{i=1}^n E_{edge}^i(r_c) }
The last sum in this expression corresponds to finite extra contributions coming from the edges

\eqn\general{E_{edge}^i={\sqrt{\lambda} \over 2\pi}\int_0^1 {ds \over s} \log \left(\frac{r_c(s)r_c(1-s)}{r_c(0) r_c(1)} \right) }
where $s$ running from zero to one parametrizes the $ith$ edge, namely $x^\mu (s)=x^\mu_i+s(x^\mu_{i+1}-x^\mu_i)$ and $r_c(s)$ is a shorthand notation for $r_c(x(s))$. For instance, a simple example is that of a cut-off that takes the value $r_c(x_i)$ at the $ith$ cusp and varies linearly between cusp and cusp, in this case

\eqn\simp{E_{edge}^i ={\sqrt{\lambda} \over 4 \pi} \log^2{r_c(x_i) \over r_c(x_{i+1})} }

\subsec{Conformal Ward identity \foot{The original idea leading to the argument below is due to Juan Maldacena.}}

An important ingredient of the computation
for the case $n=4$ was the existence of a dual $SO(2,4)$ symmetry.
 This symmetry allowed the construction of new solutions and
 fixed somehow the finite piece of the scattering amplitude.
 Naively, this conformal symmetry would imply that the amplitude is independent of
 $s$ and $t$, since they can be sent to arbitrary values by a dual conformal symmetry.
 The whole dependence on $s$ and $t$ arises due to the necessity of introducing an
 IR regulator. However, we will see that, after keeping track of the dependence on the
 IR regulator, the amplitude is still determined by the dual conformal symmetry.

To a symmetry we associate a Ward identity, that will impose certain constraints on scattering amplitudes. In order to understand these constraints for the case of the dual $SO(2,4)$ symmetry it is convenient to consider the amplitude regulated by a radial cut-off.

Given the momenta $k_i$ of the external gluons, the boundary of
the world-sheet contains cusps located at $x_i$, with $2\pi
k_i=x_i-x_{i+1}$. Now imagine that we regularize the area by
choosing a cut-off $r_c$. Moreover, we would like this cut-off to
depend on the point at the boundary we are approaching, {\it i.e.}
$r_c \rightarrow r_c(x)$. From the discussion above we expect the
regulated area to have the general form

\eqn\areg{A^{reg}_n=f \sum_{i=1}^n \log^2 \left(\frac{r_c^2(x_i)}{-2 x_{i-1,i+1}^2} \right)+Fin(x_i)}
where we have disregarded the extra terms coming from the edges since they can be seen not to affect the following argument \foot{One can convince oneself, for instance, by considering the simplified case \simp\  and applying the generator of special conformal transformations to such extra terms. It is  also instructive to apply the generator of dual conformal transformations, whose relevant piece is of the form $\int ds x^\mu(s) r_c(s)\frac{\delta}{\delta r_c(s)}$, to the general extra terms \general\ and compare this expression to eq. (34) of  \Druthree\ .} . $SO(2,4)$ transformations will then act on the points $x_i$ and $r_c(x_i)$. By requiring the area to be invariant under the action of special conformal transformations

\eqn\Konareg{K^\mu A^{reg}_n=\left(\sum_{i=1}^n 2 x_i^\mu(x_i . \partial_{x_i}+r(x_i) \partial_{r(x_i)})-x_i^2 \partial_{x_i^\mu} \right) A^{reg}_n=0}
we get an equation for the finite piece of the amplitude. This same equation was obtained perturbatively to all lops in relation with expectation values of Wilson loops in  \Druthree and at strong coupling by using dimensional regularization in \KomargodskiWA .

Supposing that the dual conformal symmetry is present beyond the
strong coupling limit, a similar argument can be extended to all
values of the coupling, {\it e.g.} by using as a regulator an
energy scale $\mu(x)$ and assuming that the amplitude has
divergences which depend only on $\mu(x)$ at the cusps (or assuming that special conformal transformations annihilate the extra pieces coming from the edges, as it happened at strong coupling.)

It turns out, that for the case of $n=4$ and $n=5$, this equation
fixes uniquely the form of the finite piece, to be the one in the
 BDS conjecture. At this point we do not know if the dual conformal symmetry
 is an exact property of planar amplitudes. We do know, however, that it is a
 symmetry of all the weak and strong coupling computations that have been
 done so far. If we assumed that it is a symmetry, then we conclude that 
the BDS conjecture for
four and five gluons is correct.

\newsec{Scattering amplitudes vs. Wilson loops}

The computation of the previous section shows a relation between two
seemingly different quantities, scattering amplitudes and expectation values of Wilson loop.

More precisely, one can consider the planar, MHV amplitude for $n$ gluons,
$A(k_1,...,k_n)$ and an associated Wilson loops in position space,
$W(x_1,...,x_n)$ formed by light-like segments joining cusps at $x_i$,
with  $2\pi k_i=x_i-x_{i+1}$.
The results of the previous section imply
that both quantities are equivalent at strong coupling.
Quite remarkable, explicit computations \Druone\BrandhuberYX\
show that this duality continues to hold also at weak coupling!
The duality between amplitudes and Wilson loops would imply the dual conformal symmetry,
since the dual conformal invariance becomes the ordinary conformal invariance of
the Wilson loop computation.

Beyond explicit computations at one loop for any number of gluons
and two loops computations to be mentioned later, expectation
values of Wilson loops were shown to posses the dual conformal
symmetry and to satisfy the same conformal ward identity \Konareg\
\Drutwo\Druthree . Such dual conformal symmetry was also observed
in perturbative scattering amplitudes \bds \BernEW \CachazoAZ\
(though it is not a proven symmetry).

The equivalence of both quantities at one loop can be stated as
follows. The BDS ansatz (which to one loop is correct by
construction) can be written as

\eqn\logA{\log M_n=Div_n+\frac{f(\lambda)}{4}a_1(k_1,...,k_n)+h(\lambda)+n k(\lambda)}
with $a_1$ the one loop amplitude and $h(\lambda)$ and
$k(\lambda)$ functions that are independent on the kinematics and
the number of gluons and we are not interested on them. On the
other hand, we can compute the one-loop expectation value of the
associated Wilson loop

 \eqn\vevWilson{< W_n >=\tilde{Div}_n+w_1(k_1,...,k_n)+c(\lambda)+n d(\lambda)}
where $c(\lambda)$ and $d(\lambda)$ are functions that are
independent on the kinematics or the number of edges and are not
interesting for us. Then, explicit computations show that
$a_1=w_1$.

Summarizing what we have said up to now:

\item{$\bullet$} The BDS ansatz implies that
the strong coupling limit of MHV planar scattering amplitudes is given by the one loop amplitude times the strong coupling limit of the cusp anomalous, that is $a_{strong}(k_1,...,k_n)=f^{strong}a_1(k_1,...,k_n)$.

\item{$\bullet$} Our computation implies that scattering amplitudes and expectation values of Wilson loops agree at strong coupling, namely $a_{strong}(k_1,...,k_n)=w_{strong}(k_1,...,k_n)$.

\item{$\bullet$}  While explicit one loop computations show that $a_1(k_1,...,k_n)=w_1(k_1,...,k_n)$.

Assuming BDS and using the next two results we arrive to

\eqn\bdstest{w_{strong}(k_1,...,k_n)=f^{strong}w_1(k_1,...,k_n)}

So, the expectation value of a Wilson loop at strong coupling is
given by its one loop expectation value times the cusp anomalous
dimension at strong coupling. Obviously, that is a very non
trivial statement, and indeed, for the case of four and five
edges, that is the case! However, as we have seen, in this case
the expectation value is fixed by symmetries. Hence, in order to
test the BDS conjecture, we need to consider polygons with more
than five edges.

\subsec{Testing BDS}

As just mentioned, in order to test the BDS conjecture,
one would need to consider polygons of more than five edges. It seems
very difficult to find explicit solutions for six edges.
However,  we consider a zig-zag configuration with a large number of edges
 that approximates the rectangular Wilson loop.

\ifig\rwilson{Zig-Zag configuration approaching the space-like
rectangular Wilson loop. } {\epsfxsize2.5in\epsfbox{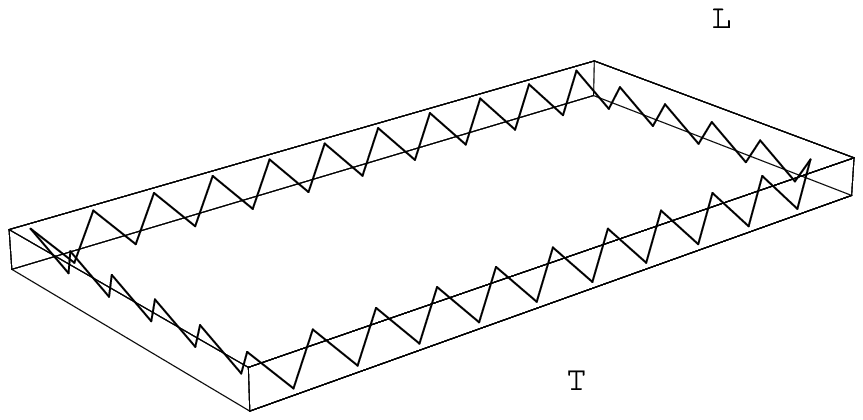}}

In the limit of very large $T$ and $L$ and for $T \gg L$, we can compute the expectation value both at weak and strong coupling \ReyIK\MaldacenaIM , obtaining

\eqn\rectWl{\log <W_{rect}^{weak}> =\frac{\lambda}{8 \pi} \frac{T}{L},~~~~~\log <W_{rect}^{strong}> =\frac{\sqrt{\lambda} 4 \pi^2}{\Gamma(1/4)^4} \frac{T}{L}}
While the BDS ansatz prediction would be $\log <W_{rect}^{strong}> =\frac{\sqrt{\lambda} }{4} \frac{T}{L}$. Hence, the BDS ansatz needs to be revised for a large number of gluons.

The previous reasoning can be also repeated by considering the two loops amplitude
 versus the two loops Wilson loop expectation values.
  Explicit two loops computations for the rectangular Wilson loop,
   show that at this order and for a large number of gluons,
   either BDS or the duality between scattering amplitudes and Wilson loops fails.

\subsec{$n=6$ case}

As explained above, the presence of the dual conformal symmetry
fixes both the scattering amplitudes and the expectation values of
the Wilson loops for the case of $n=4$ and $n=5$, and in both
cases, the result agrees with the BDS ansatz.

The BDS ansatz for six gluons satisfies the dual conformal Ward identities, however, it is not uniquely fixed by these. The general solution is the BDS ansatz plus an arbitrary function of invariant cross ratios

\eqn\cross{\eqalign{K.f(u_1,u_2,u_3)=0~~~\rightarrow~~~A_6=A_{BDS}+f(u_1,u_2,u_3),\cr u_1=\frac{x_{13}^2 x_{46}^2}{x_{14}^2 x_{36}^2},~~u_2=\frac{x_{24}^2 x_{15}^2}{x_{25}^2 x_{14}^2},~~u_3=\frac{x_{35}^2 x_{26}^2}{x_{36}^2 x_{25}^2}}}
Note that this invariant cross ratios cannot be constructed for
$n<6$. A remarkable explicit computation for the scattering of six
gluons at two loops \BernAP, shows that indeed $f \neq 0$ and
hence the BDS ansatz is to be modified for six gluons at two
loops.

A parallel computation for the two loops expectation value of the
associated Wilson loops has also been carried out \Drufour . Quite
remarkably, the duality between scattering amplitudes and Wilson
loops continues to hold for this case! \Drufive\BernAP . This is a
strong indication that the duality may be true for any number of
gluons at any loop order.

\newsec{Conclusions}

In this notes we have described recent progress in computing
planar scattering amplitudes on ${\cal N}=4$ SYM at strong
coupling by using the AdS/CFT correspondence. The computation
reduces to a minimal surface problem in $AdS$, with boundary
conditions fixed by the momenta of the external particles.

Amplitudes are IR divergent and a regulator needs to be introduced
in order to define them properly. We perform explicit computations
both by using dimensional regularization and a cut-off in the
radial direction.  While the former scheme allows a direct
comparison with gauge theory results, the former is more
convenient for understanding the conformal properties of the
amplitudes.

One of our main motivations was to test the BDS ansatz. Our
results agree with this conjecture for $n=4,5$ but disagree for a
large number of gluons. The agreement can be understood as due to
the dual conformal symmetry.
Moreover, explicit computations indeed show that the
BDS ansatz is not correct for six gluons at two loops.

An important ingredient in the computation of amplitudes at strong coupling is the presence of a dual $SO(2,4)$ conformal
symmetry. We presented a simple argument leading to a Ward
identity for this symmetry.  In addition, The strong coupling picture suggests a relation between amplitudes
and Wilson loops, which seems to be a true relation and it
survived the nontrivial check of  \Drufive\BernAP.

\bigskip

There are many directions one could try to follow

\bigskip

\item{$\bullet$}  Construct new solutions corresponding to the scattering of more than four
 gluons. Despite some partial progress \JevickiPK\nastone\Moroone\Morotwo ,
 general
 solutions other than the one for $n=4$ are missing. Such solutions would be very useful in trying to understand the existence of iterative relations, from the strong coupling side.

\item{$\bullet$}  Try to use the machinery of integrability in order to find new solutions, or the value of the action even without knowing the classical solutions. Besides, integrability may provide a set of constraints that would fix the form of the amplitudes.

\item{$\bullet$}  Compute subleading corrections in $1/\sqrt{\lambda}$. Among other things, one should be able to compute the dependence on the helicities of the particles and understand the particular role played by MHV amplitudes. Some attempts have been done in \KruczenskiCY , where apparently is subtle to extend dimensional regularization beyond the classical analysis. Maybe it is convenient to consider other schemes, like a radial cut-off for instance.

\item{$\bullet$} Try to understand higher genus corrections. These would correspond to non planar corrections to scattering amplitudes.

\item{$\bullet$} The extension of the prescription described here to other backgrounds, less super-symmetryc or without conformal invariance, is also an important problem. See \KomargodskiER\McGreevyKT\ItoZY\OzQR for recent interesting developments in this direction.

\item{$\bullet$} 

\item{$\bullet$}  An interesting problem would be to try to determine the duality between MHV scattering amplitudes and expectation values of Wilson loops. In case this duality holds true, it would be interesting to extend this equivalence to non MHV amplitudes.

\item{$\bullet$} The apparent equivalence between scattering
amplitudes and Wilson loops for the case of six legs, hints to the
existence of new symmetries that would fix the form of both
quantities in this case.

\item{$\bullet$} Finally, it would be very interesting to find appropriate
modifications to the BDS ansatz,
in order to describe higher point amplitudes at all values of the coupling. Though it would
be very surprising to find a general explicit formula.

{\bf Acknowledgments}

These lectures are based on results obtained in collaboration with Juan Maldacena. I would like to thank  G. Korchemsky, A. Murugan, A. Sever, E. Sokatchev and particularly Juan Maldacena for enlightening discussions and the hospitality of the many institutions in which I presented the work described in these notes. The work of the author was supported by VENI grant 680-47-113.

\listrefs

\bye